 \documentclass[preprint2]{aastex}

\def\dsm{$\mathrm{M}_\odot$}

\shorttitle{Evolution of rotational velocities of A-type stars}
\shortauthors{Yang et al.}

\begin{document}


\title{Evolution of rotational velocities of A-type stars}

\author{Wuming Yang\altaffilmark{1,2}, Shaolan Bi\altaffilmark{1},
Xiangcun Meng\altaffilmark{2}, Zhijia Tian\altaffilmark{1}}

\altaffiltext{1}{Department of Astronomy, Beijing Normal University,
Beijing 100875, China; yangwuming@ynao.ac.cn; yangwuming@bnu.edu.cn}
\altaffiltext{2}{School of Physics and Chemistry, Henan Polytechnic
University, Jiaozuo 454000, Henan, China}

\begin{abstract}
It was found that the equatorial velocity of A-type stars undergoes an acceleration in the
first third of the main sequence (MS) stage, but the velocity decreases as if the stars were
not undergoing any redistribution of angular momentum in the external layers in the last
stage of the MS phase. Our calculations show that the acceleration and the decrease of
the equatorial velocity can be reproduced by the evolution of \textbf{the differential 
rotation zero-age MS model} with the angular momentum transport caused by hydrodynamic instabilities
during the MS stage. The acceleration results from the fact that the angular momentum
stored in the interiors of the stars is transported outwards. In the last stage,
the core and the radiative envelope are uncoupling, and the rotation of the envelope is
a quasi-solid rotation; the uncoupling and the expansion of the envelope lead to that
the decrease of the equatorial velocity approximately follows the slope for the change
in the equatorial velocity of the model without any redistribution of angular momentum.
When the fractional age 0.3 $\lesssim\mathrm{t/t_{MS}}\lesssim$ 0.5, the equatorial
velocity remains almost constant for the stars whose central density increases with
age in the early stage of the MS phase, while the velocity decreases with age for
the stars whose central density decreases with age in the early stage of the MS phase.
\end{abstract}


\section{Introduction}
Helioseismology and asteroseismology have successfully shown us the internal
rotation of the Sun and red-giant stars. For example, the Sun has a flat
rotational profile \citep{brow89, koso97} but red-giant stars have a fast
rotation core \citep{aert03, beck12}, which provided some constraints on
the processes of angular momentum transport in the Sun and red-giant stars.
In order to reproduce the flat solar rotational profile, the magnetic angular
momentum transport \citep{egge05, yang06} or the gravity-wave angular momentum
transport \citep{zahn97, talo05} should be considered in solar models,
besides the angular momentum transport caused by hydrodynamic instabilities.

Most of A-type stars are known to be fast rotators, with a typical rotational velocity
of about 160 km $\mathrm{s}^{-1}$ \citep{roye07} for stars with masses between about
1.3 and 3.0 \dsm{}. Thus there may be a non-negligible effect of rotation on
the structure and evolution of these stars. However, these stars are always hotter than
the red edge of the instability strip \citep{saio98}. They are generally not
expected to exhibit solar-like oscillations. Therefore the effect of rotation and rotation
scenarios of these stars are hard to be understood as that of the Sun and red-giant stars
via seismology.

Fortunately, the surface rotational velocities of thousands of stars with the spectral type
between late B and early F have been observed \citep{abt95, wolf97, roye02a, roye02b, wolf04,
roye04a, roye04b, roye07, diaz11, zore12}. These stars include pre-main sequence (PMS)
and main sequence (MS) stars. The observed surface rotational velocity of a star depends on:
(1) the initial angular momentum, (2) the rate of angular momentum loss, (3) the angular momentum
transport in stellar interiors, and (4) the changes in the moment of inertia of the star.
For stars with masses greater than 1.6 \dsm{}, there is no firm evidence of activity and
any significant angular momentum loss during MS evolution \citep{wolf97}. For these
MS stars, thus, the observed surface rotational velocities are mainly dependent on
the initial rotational profile at the zero-age main sequence (ZAMS),
the angular momentum transport in stellar interiors, and the changes in the moment
of inertia of stars. The changes of the moment of inertia can be understood
from evolutionary models. Therefore, the observed rotational velocities of A-type stars could
provide important clues about the initial rotational profile at the ZAMS and the efficiency of
angular momentum transport in stellar interiors.

Recently, \cite{zore12} studied the evolution of surface rotational velocities of more than
one thousand A-type MS stars. They found that the surface velocities of the stars with
masses between about 1.7 and 3.2 \dsm{} undergo a strong acceleration in the first third of the
MS evolutionary phase, which strongly differs from that theoretically predicted by two
limiting cases of internal angular momentum redistributions: (i) rigid rotation;
(ii) conservation of angular momentum in stellar shells, while in the last third
of the MS the velocities decrease as if the stars were not undergoing any redistribution
of the angular momentum in the external layers \citep{zore12}.

The ZAMS models are always assumed to be a uniform rotation in the calculations
of the evolution of rotating stars. If this assumption was correct, a special mechanism
might be required to extract angular momentum from the stellar interiors to achieve
the acceleration of the equatorial velocity. In fact, \cite{wolf97} and \cite{wolf04}
had studied the angular momentum evolution of A- and F-type stars from the birth line
to the MS. They concluded that A- and early F-type stars do not loss angular momentum
and the angular momentum could be conserved in the shells of the stars as the stars
evolve from the end of fully convective phase to the ZAMS. Thus, for these stars the rotation
at the ZAMS should be differential \textbf{\textit{(hereafter this differential rotation
ZAMS model obtained from the evolution with the conservation of angular momentum in shells
was referred to Wolff ZAMS model)}}, and the core could serve as a reservoir of angular
momentum which can be transported from the fast rotation core to the slow rotation envelope to
produce the acceleration of the equatorial velocity. Moreover, the too high or too low efficiency
of angular momentum transport in stellar interiors could not reproduce the acceleration of
the surface velocity in the first third of the MS evolutionary phase \citep{zore12}.
Thus the results of \cite{zore12} provide us an opportunity to test the conclusions
of \cite{wolf04} and the angular momentum transport in the interiors of A-type stars.

In this work, we focus mainly on the evolution of the equatorial velocity of
A-type stars. The paper is organized as follow. We simply show our stellar
models in section 2. The results are represented in section 3. In section 4,
we discuss and summarize the results.

\section{Stellar models and calculation results}
\subsection{Stellar models}

We used the Yale Rotation Evolution Code \citep{pins89, yang07} to compute
the evolutions of rotating models with \textbf{different masses and metallicities}.
Hydrodynamical instabilities considered in this code are meridional circulation,
the Goldreich-Schubert-Fricke
instability, and the secular shear instability \citep{enda78, pins89}.
The OPAL EOS tables \citep{roge02}, OPAL opacity tables \citep{igle96},
and the opacity tables for low temperature provided by \cite{alex94} were used.
Energy transfer by convection is treated according to the standard mixing length theory.
The value of 1.72 for the mixing-length parameter ($\alpha$) was calibrated against the Sun.
\textbf{The initial metallicity $Z$ was fixed at 0.02 and 0.008, and the hydrogen abundance
was determined by $X=0.767-3Z$.} For a given mass, the initial angular momentum is estimated by
using the formula of \cite{kawa87}. Angular momentum loss due to magnetic braking
and mass loss could be negligible in A-type stars \citep{macg94, wolf97, zore12}.
Thus we assumed that the total angular momentum is conserved in our models.

We calculated the following four evolutionary cases: (1) C1, models were evolved
from the PMS to the terminal age main sequence (TAMS) without any exchange of angular
momentum in shells; (2) C2, models were computed as a rigid rotator; (3) C3; and (4)
C4. In the last two cases, models were evolved from the ZAMS to the TAMS with angular
momentum transport caused by the rotationally induced instabilities, and the ZAMS
model is a uniform rotator for the C3 but is a \textbf{Wolff ZAMS model} for the C4.
Moreover, we assumed that the rotation of the surface layers
(the fractional mass $\delta M/M \simeq 10^{-4}$) is uniform.

\subsection{Calculation results}

\begin{figure}
\includegraphics[angle=-90, scale=0.6]{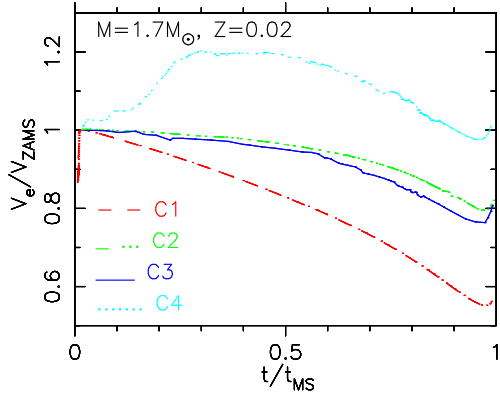}
\caption{Evolution of the equatorial velocity in the MS stage. The long-dashed (red) line
shows the evolution of C1. The dash-triple-dotted (green) line indicates the result of C2.
The solid (blue) line is given for C3. The dotted (cyan) line corresponds to the evolution
of C4. \label{f17}}
\end{figure}

Figure \ref{f17} shows the evolution of the equatorial velocity of models with $M=$ 1.7 \dsm{}
and $Z=0.02$ in the MS life time span ($\mathrm{t_{MS}}$). The notation $\mathrm{V_{ZAMS}}$ is
the equatorial velocity at the ZAMS. For the cases C1, C2, and C3, from the ZAMS to
$\mathrm{t/t_{ZAMS}} \approx$ 0.95, the equatorial velocities decrease. In case C1,
the equatorial velocity descends about \textbf{45}\%, which is consistent with the expectation
of the law of conservation of angular momentum in shells ($V_{e}=V_{0}R_{0}/R_{e}$,
where $R_{e}$ is the radius of stars and $V_{0}$ and $R_{0}$ refer to initial values).
In the case C2, the equatorial velocity decreases around 20\%. These two limiting cases
did not reproduce the acceleration of the surface velocity in the early stage of MS.
The evolution of the equatorial velocity of case C3 is similar to that of the case C2
except for in the last stage of the MS. This is because the rotation of the radiative
region of the MS models of the C3 is a quasi-solid body rotation and the core and
the envelope are uncoupling in the last stage. In the case C4, the equatorial velocity
undergoes an acceleration in the early stage of the MS phase.
From $\mathrm{t/t_{MS}}\simeq$ 0.017 to $\mathrm{t/t_{MS}}\simeq$ 0.3, the equatorial velocity
increases about 20\%. But when $\mathrm{t/t_{MS}}$ increases from about 0.3 to 0.5,
the velocity remains almost constant. After $\mathrm{t/t_{MS}} >$ 0.5,
the velocity decreases as evolution proceeds until the evolution
approaches the end of the MS stage. When the fractional age $\mathrm{t/t_{MS}} >$ 0.7,
the decrease of the equatorial velocity of the case C4 is similar to that
of the case C1, which is consistent with the finding of \cite{zore12}.

\begin{figure}
\includegraphics[angle=-90, scale=0.5]{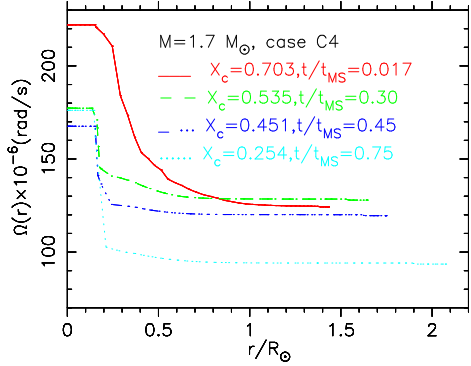}
\caption{The radial distributions of internal angular velocity at different evolutionary stages.
The solid (red) line shows the distribution at $X_{c}=0.703$. The long-dashed (green) line
indicates the distribution at $X_{c}=0.535$. The dash-triple-dotted
(blue) line shows the distribution at $X_{c}=0.451$, and the dotted (cyan)
line refers to the distribution at $X_{c}=0.254$. \label{fom1}}
\end{figure}

\begin{figure}
\includegraphics[angle=-90, scale=0.5]{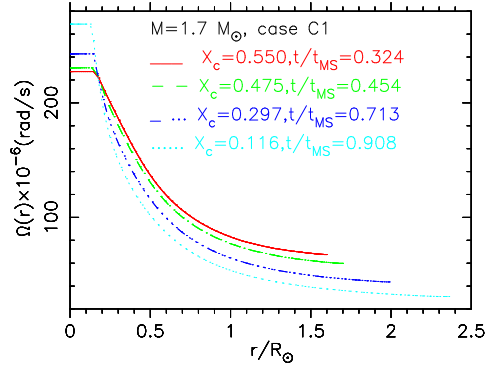}
\caption{Same as Fig. \ref{fom1} but for the evolutions of case C1. \label{fom}}
\end{figure}

Figure \ref{fom1} shows the radial distributions of internal angular velocity of the model
with $M=$ 1.7 \dsm{} at different stages of the evolution of case C4.
Because there is no redistribution of angular momentum in shells during the PMS stage and
the core of the model contracts but the envelope expands, the core rotates faster
than the envelope when the model evolves to the ZAMS. From $\mathrm{t/t_{MS}}\simeq$ 0.017 to
$\mathrm{t/t_{MS}}\simeq$ 0.30, the angular momentum stored in the fast rotation
core is transported to the slow rotation envelope, which almost fully impedes the decrease in
the angular velocity of the envelope, even leads to an increase in the angular velocity.
In addition, due to the increase in radius, the equatorial velocity increases obviously.
When the model evolves from $\mathrm{t/t_{MS}} \approx$ 0.3 to $\mathrm{t/t_{MS}} \approx$ 0.5,
the core and the envelope are strongly coupled by hydrodynamic instabilities. Although
the core still contracts, its angular velocity decreases with the decrease in the angular velocity
of the envelope. The angular momentum stored in the stellar interiors is continually transported
outwards, which is insufficient to remain the acceleration of the equatorial velocity but
can allow the velocity to remain almost constant. When $\mathrm{t/t_{MS}} >$ 0.6,
due to the facts that the core contracts and the envelope
expands rapidly, the hydrodynamical instabilities are insufficient to couple the core to
the envelope. Thus the core and the envelope are uncoupling. However, the instabilities are
sufficient to keep the envelope rotating as a quasi-solid body. In our models, the
differential rotation results from the contraction/expansion of stars.
Due to the uncoupling of the core and the envelope, the decrease of the angular
velocity of the envelope depends almost only on the expansion of the envelope. Moreover,
Fig. \ref{fom} shows that most of the radiative region spin down at an approximately equal rate
from $\mathrm{t/t_{MS}}\approx$ 0.7 to $\mathrm{t/t_{MS}} \approx$ 0.9 in the evolution of case C1,
which is similar to the manner of quasi-solid rotation.
Thus when the fractional age $\mathrm{t/t_{MS}} >$ 0.7, the decrease in the equatorial
velocities of cases C1 and C4 follows an approximate slope.

\begin{figure}

\includegraphics[angle=-90, scale=0.4]{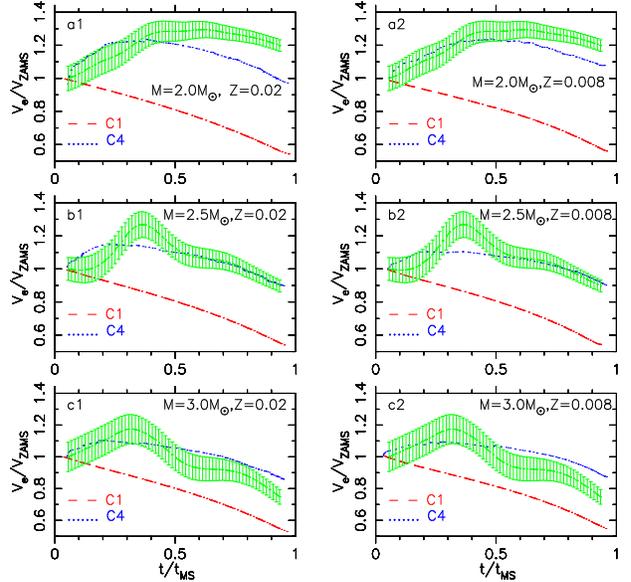}
\caption{Comparisons of the evolutions of theoretically equatorial velocities
with the observed data. The dash-dotted lines (green) show the results of \cite{zore12}.
The error bars (green) represent $1\sigma$ errors.
\label{fcomp}}
\end{figure}

\textbf{We also calculated the evolutions of the equatorial velocity of models with
$M=$ 2.0, 2.5, and 3.0 \dsm{}. The comparisons of the evolutions of the equatorial
velocities of our models with the results of \cite{zore12} are shown in Fig. \ref{fcomp}.}
The evolutions of cases C1, C2, and C3 of these models are similar to those of model
with $M=$ 1.7 \dsm{}, \textbf{i.e. the equatorial velocity decreases with age.
But} in the early stage ($\mathrm{t/t_{MS}} \lesssim$ 0.3) of the MS of the case C4,
the equatorial velocity undergoes an acceleration.
However, \textbf{the magnitude of} the increase of the ratio $\mathrm{V_{e}/V_{ZAMS}}$ of
\textbf{our models} is obviously lower than that \textbf{obtained by \cite{zore12}.
For example, at $\mathrm{t/t_{MS}} = 0.35$, the value of the observed $\mathrm{V_{e}/V_{ZAMS}}$
is 1.265 $\pm$ 0.079 for the star with $M=$ 2.5 \dsm{}, but the value for our models is about
1.11. The discrepancy is about $2\sigma$}. When the fractional age $\mathrm{t/t_{MS}}$
increases from around 0.3 to 0.5, \textbf{the theoretical velocities decrease slightly
but the observed ones decrease rapidly for stars with $M=$ 2.5 and 3.0 \dsm{}. When the age
$\mathrm{t/t_{MS}}> 0.5$, the value of the theoretical $\mathrm{V_{e}/V_{ZAMS}}$ is lower
for the star with $M=2.0$ \dsm{} but is higher for the star with $M=3.0$ \dsm{}
than the observed one. However, for the star with $M=2.5$ \dsm{}, the evolution of the
theoretical $\mathrm{V_{e}/V_{ZAMS}}$ is good agreement with the observed one}.
After $\mathrm{t/t_{MS}} >$ 0.7, the decrease of the velocities of cases C1 and C4
\textbf{and the observed one} follow an approximate slope.

\textbf{Furthermore, the Fig. \ref{fcomp} shows that the evolution of the equatorial
velocities is slightly affected by the metallicity. This can be due to the fact that
the radius of low-Z model is smaller than that of high-Z model.}

\section{Discussion and Conclusions}

\begin{figure}
\includegraphics[angle=-90, scale=0.6]{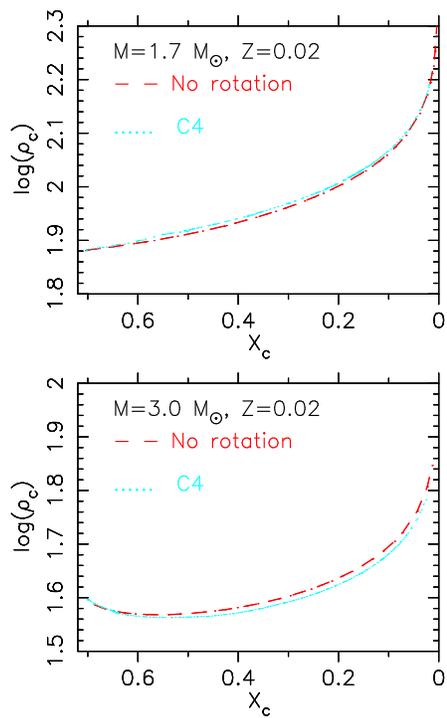}
\caption{The central density ($\rho_{c}$) as a function of the mass fraction of central hydrogen.
The dashed (red) lines show the results of models without rotation.
The dotted (cyan) lines indicate the results of models with rotation. \label{frc}}
\end{figure}

The increase of the fractional velocity $\mathrm{V_{e}/V_{ZAMS}}$ is mainly dependent on the
angular momentum that is stored in the interiors of the star and can be transported
outwards to accelerate the surface layers. The angular momentum depends on the contraction
of the stellar core and the expansion of the envelope of the star in our models.
The more massive the star, the less the contraction of its core when the star evolves
from the PMS to the ZAMS. In addition, in the early stage of the MS, the central density
increases as the evolution proceeds for the stars with $M \lesssim$ 2.0 \dsm{}, but
the central density decreases (see the Fig. \ref{frc}) for the stars with $M \gtrsim$ 2.1 \dsm{},
i.e. the angular velocity of the core decreases for stars with $M \gtrsim$ 2.1 \dsm{}.
As a consequence, the more massive the star, the smaller the fractional angular momentum
that can be transported outwards. Moreover, the more massive the star, the larger
the expansion of the radius during the MS stage, thus the more difficult to remain
the angular velocity of the surface layer. Therefore the increase of the fractional
velocity decreases with increasing the mass, and the available angular momentum stored
in the stellar interiors is almost drained when $\mathrm{t/t_{MS}} \approx$ 0.3
for the stars with $M >$ 2.1 \dsm{}. Hence from $\mathrm{t/t_{MS}} \approx$ 0.3 to
$\mathrm{t/t_{MS}} \approx$ 0.5, the equatorial velocity decreases for stars
with $M >$ 2.1 \dsm{}.

\textbf{The instabilities caused that the angular momentum was transported outwards in the early
stage of the MS, which leads to the increase in the $\mathrm{V_{e}/V_{ZAMS}}$.}
Although the magnitude of the increase in the fractional velocity of our models is less
than that obtained by \cite{zore12}, who given the magnitude as high as 30\%, the velocity
increases as the evolution proceeds during the first third of the MS phase, which is
consistent with the finding of \cite{zore12}. This might offer a support for \textbf{the Wolff
ZAMS model and imply that the core and the envelope is uncoupling as A-type stars evolve from
the end of fully convective phase to the ZAMS but is coupling in the early stage of the MS.
The theoretical magnitude is lower than the observed one. This might be caused by that
the efficiency of angular momentum transport wasn't calibrated.}

In the last stage of the MS (0.7 $\lesssim \mathrm{t/t_{MS}} \lesssim$ 0.95),
the core and the radiative envelope are decoupling in the evolution of cases C1,
C3, and C4. The contraction of the core almost not affect the rotation of the envelope.
For the case C1, although the rotation of the whole radiative envelope is differential,
the envelope spins down at an approximately equal rate in our models, which is similar to the manner
of the quasi-solid body rotation. For the case C4, the rotation of the radiative envelope
is a quasi-solid rotation. Thus the decrease of the fractional velocity of the cases C1 and C4
follows an approximate slope when $\mathrm{t/t_{MS}} \gtrsim$ 0.7. As a consequence,
although the velocities of A-type stars decrease as if the stars were not undergoing
any redistribution of the angular momentum in their envelope during the final stages of
the MS phase, the rotation of the envelope of A-type stars might be a quasi-solid rotation
with the decoupling of the core and the envelope. \textbf{Thus, the rotation of A-type stars
might be closer to the quasi-solid body rotation during $\mathrm{t/t_{MS}} \approx$ 0.3 - 0.5
than in the early or last stage of the MS. }

\textbf{Compared with \cite{ekst08} models used by \cite{zore12}, our ZAMS models for C4 are
the Wolff ZAMS model but the rotation of the ZAMS models of \cite{ekst08} and our C3 is
a solid body rotation. The equatorial velocity of our C3 decreases with age,
which is similar to that obtained by \cite{zore12} from \cite{ekst08} models.
The cases C3 and C4 have the same instabilities, this may imply that the Wolff
ZAMS model is required to achieve the acceleration.
Moreover, we neglected the angular momentum loss caused by mass loss. Finally,
there are some differences in the treatments of instabilities included models
\citep{pins89, maed98}, which can lead to the differences in the efficiency of angular momentum
transport and the mixing of elements. Besides the hydrostatic and Von Zeipel effects,
the mixing caused by rotationally induced instabilities is one of the key factors that
affects the distribution of density and the instability of convection by changing
the distribution of elements. In our models, the effects of rotation lead to a decrease
in the convective core for stars with $M\lesssim$ 2.0 \dsm{} but an increase in the
convective core for stars with $M \gtrsim$ 2.1 \dsm{} \citep{yang12}. As a consequence,
near the end of the MS, the radius of rotating models is smaller than that of non-rotating
ones for the stars with $M\lesssim$ 2.0 \dsm{} but larger than that of non-rotating models for
the stars with $M \gtrsim$ 2.1 \dsm{}. Due to the fact that the decrease of the ratio
$\mathrm{V_{e}/V_{ZAMS}}$ is relevant to the increase in the radius, thus the instabilities
make the stars with $M\lesssim$ 2.0 \dsm{} easier remain their $\mathrm{V_{e}/V_{ZAMS}}$
at a high value than the stars with $M \gtrsim$ 2.1 \dsm{}.}

In this work, we calculated the evolution of the equatorial velocity of A-type stars
with different conditions. The evolution of the case C4 reproduces the most characteristics
of the equatorial velocity of A-type stars that were found by \cite{zore12}.
From the ZAMS to $\mathrm{t/t_{MS}} \approx$ 0.3, the equatorial velocity undergoes
an acceleration. But in the last stage of the MS stage ($\mathrm{t/t_{MS}} \gtrsim$ 0.7),
the core and the envelope of A-type stars are uncoupling; the envelope rotates as a quasi-solid body,
but the decrease of the equatorial velocity approximately follows the slope of the velocity of
the model without any redistribution of the angular momentum in this stage. However,
the magnitude of the increase of the fractional velocity $\mathrm{V_{e}/V_{ZAMS}}$ of
our models is less than that obtained by \cite{zore12}. Moreover, our models with \textbf{$M=$
2.5} and 3.0 \dsm{} did not reproduce the rapid decrease in the equatorial
velocity during the middle stage of the MS phase. \textbf{This might imply that some mechanisms
for the transport or loss of angular momentum are missing in our models.}

\acknowledgments \textbf{We thank the anonymous referee for his/her helpful comments and
J. Zorec and F. Royer for providing their data and acknowledge} the support from
the NSFC 11273012, 11273007, 10933002, and 11003003, and the Project of Science and Technology
from the Ministry of Education (211102).


\clearpage

\end{document}